\documentclass[12pt]{article}
\usepackage{graphicx}
\usepackage{amsmath,amsthm}
\usepackage{amsfonts}
\usepackage{amssymb}
\usepackage{appendix}
\usepackage{fullpage}
\usepackage{xcolor}
\definecolor{violet}{rgb}{0.1,0.1,1}
\usepackage[colorlinks]{hyperref}
\hypersetup{
  citecolor=blue,
  filecolor=orange,
  linkcolor=red
}

\newcommand{\Q}{{\mathbb Q}}
\newcommand{\C}{{\mathbb C}}
\newcommand{\Qbar}{{\kern.1ex\overline{\kern-.1ex\Q\kern-.1ex}\kern.1ex}}
\newcommand{\Fbar}{{\kern.1ex\overline{\kern-.3ex F \kern+.1ex}\kern.1ex}}

\newcommand{\R}{{\mathbb R}}
\newcommand{\Af}{{\mathbb A}}  % For Affine A
\newcommand{\Z}{{\mathbb Z}}

\newcommand{\N}{{\mathbb N}}

\newcommand{\W}{{\mathcal W}}

\newcommand{\kM}{{ (n-r)^2 }}    %% for upper bound for rigidity.
\renewcommand{\le}{\leqslant}
\renewcommand{\ge}{\geqslant}
\renewcommand{\leq}{\leqslant}
\renewcommand{\geq}{\geqslant}

\newcommand{\RIG}{{\sf RIG}}  %This is a set
\newcommand{\RANK}{{\sf RANK}} % This is a set

\DeclareMathOperator{\rk}{rank}
\DeclareMathOperator{\rank}{rank}
\DeclareMathOperator{\Rig}{Rig}
\DeclareMathOperator{\trdeg}{tr\ deg}
\DeclareMathOperator{\minors}{Minors}
\DeclareMathOperator{\supp}{Supp}
\DeclareMathOperator{\image}{Im}

\newtheorem{theorem}{Theorem}
\newtheorem{lemma}[theorem]{Lemma}
\newtheorem{claim}[theorem]{Claim}
\newtheorem{proposition}[theorem]{Proposition}

\newtheorem{corollary}[theorem]{Corollary}
\newtheorem{definition}[theorem]{Definition}

\newtheorem{remark}[theorem]{Remark}

\newcommand{\nbyn}{\ensuremath n \times n}

\begin{document}
\title{Using Elimination Theory to construct Rigid Matrices} 
\author{
  Abhinav Kumar\thanks{{\tt abhinav@math.mit.edu}, Department of
    Mathematics, Massachusetts Institute of Technology, Cambridge,
    USA. This work was started when the author was with Microsoft
    Research, Redmond, and later supported by NSF CAREER grant
    DMS-0952486.},
  ~~Satyanarayana V. Lokam\thanks{{\tt satya@microsoft.com}, Microsoft
    Research India, Bangalore, India.},\\[1mm]
  Vijay M. Patankar\thanks{{\tt vijay@isichennai.res.in}, Indian
    Statistical Institute, Chennai Centre, Chennai, India. This work
    was started and major part of this work was completed while the
    author was with Microsoft Research India, Bangalore.},
  ~~Jayalal Sarma M.N.\thanks{{\tt jayalal@cse.iitm.ac.in}, Department
    of Computer Science \& Engineering, Indian Institute of Technology
    Madras, Chennai, India. Work done while the author was with the
    Institute of Mathematical Sciences, Chennai, and Institute for
    Theoretical Computer Science, Tsinghua University, Beijing,
    China.}}

\maketitle

\begin{abstract}
The rigidity of a matrix $A$ for target rank $r$ is the minimum number
of entries of $A$ that must be changed to ensure that the rank of
the altered matrix is at most $r$.  Since its introduction by Valiant
\cite{Val77}, rigidity and similar rank-robustness functions of
matrices have found numerous applications in circuit complexity,
communication complexity, and learning complexity. Almost all $\nbyn$
matrices over an infinite field have a rigidity of $(n-r)^2$. It is a
long-standing open question to construct infinite families of
\emph{explicit} matrices even with superlinear rigidity when $r=\Omega(n)$.

In this paper, we construct an infinite family of complex matrices
with the largest possible, i.e., $(n-r)^2$, rigidity. The entries of
an $\nbyn$ matrix in this family are distinct primitive roots of unity
of orders roughly {$\exp(n^2 \log n)$}. To the best of our knowledge, this is
the first family of concrete (but not entirely explicit) matrices
having maximal rigidity and a succinct algebraic description.

Our construction is based on elimination theory of polynomial
ideals. In particular, we use results on the existence of polynomials
in elimination ideals with effective degree upper bounds (effective
Nullstellensatz). Using elementary algebraic geometry, we prove that
the dimension of the affine variety of matrices of rigidity at
most $k$ is exactly $n^2 - (n-r)^2 +k$. Finally, we use elimination theory to
examine whether the rigidity function is semicontinuous.
\end{abstract}

\section{Introduction}
\label{sec:intro}

Valiant \cite{Val77} introduced the notion of matrix rigidity. The
rigidity function $\Rig(A,r)$ of a matrix $A$ for target rank $r$ is
defined to be the smallest number of entries of $A$ that must be
changed to ensure that the altered matrix has rank at most $r$. It is
easy to see that for every $\nbyn$ matrix $A$ (over any field),
$\Rig(A,r) \le (n-r)^2$. Valiant also showed that, over an infinite
field, almost all matrices have rigidity exactly $(n-r)^2$. It is a
long-standing open question to construct infinite families of
\emph{explicit} matrices with superlinear rigidity for $r =
\Omega(n)$. Here, by an explicit family, we mean that the $\nbyn$
matrix in the family is computable by a deterministic Turing machine
in time polynomial in $n$ or by a Boolean circuit of size polynomial
in $n$.  Lower bounds on rigidity of explicit matrices are motivated
by their numerous applications in complexity theory. In particular,
Valiant showed that lower bounds of the form $\Rig(A,\epsilon n) =
n^{1+\delta}$ (where $\epsilon$ and $\delta$ are some positive
constants) imply that the linear transformation defined by $A$ cannot
be computed by arithmetic circuits of linear size and logarithmic
depth consisting of gates that compute linear functions of their
inputs. Since then, applications of lower bounds on rigidity and
similar rank-robustness functions have been found in circuit
complexity, communication complexity, and learning complexity (see
\cite{FKLMSS01, For02, Raz89, Lok95, PP04, LS06}). For comprehensive
surveys on this topic, see \cite{Cod00}, \cite{Che05}, and
\cite{Lok09}.  Over finite fields, the best known lower bound for
explicit $A$ was first proved by Friedman~\cite{Fri93} and is
$\Rig(A,r) =\Omega(\frac{n^2}{r} \log \frac{n}{r})$ for parity check
matrices of good error-correcting codes. Over infinite fields, the
same lower bound was proved by Shokrollahi, Spielman, and
Stemann~\cite{SSS97} for Cauchy matrices, Discrete Fourier Transform
matrices of prime order {(see \cite{Lok00})}, and other families.
Note that this type of lower bound reduces to the trivial
$\Rig(A,r)=\Omega(n)$ when $r=\Omega(n)$. In \cite{Lok06}, lower
bounds of the form $\Rig(A,\epsilon n) = \Omega(n^2)$ were proved when
$A=(\sqrt{p_{jk}})$ or when $A = (\exp(2 \pi {\sf i}/p_{jk}))$, where
$p_{jk}$ are the first $n^2$ primes. These matrices, however, are not
explicit in the sense defined above.

In this paper, we construct an infinite family of complex matrices
with the highest possible, i.e., $(n-r)^2$, rigidity. The entries of
the $\nbyn$ matrix in this family are primitive roots of unity of
orders roughly {$\exp(n^2 \log n)$}. We show that the real parts
of these matrices are also maximally rigid. Like the matrices in
\cite{Lok06}, this family of matrices is not explicit in the sense of
efficient computability described earlier. However, one of the
motivations for studying rigidity comes from algebraic complexity. In
the world of algebraic complexity, any element of the ground field (in
our case $\C$) is considered a primitive or atomic object. In this
sense, the matrices we construct are explicitly described algebraic
entities. To the best of our knowledge, this is the first construction
giving an infinite family of non-generic/concrete matrices with
maximum rigidity. It is still unsatisfactory, though, that the roots
of unity in our matrices have orders exponential in $n$. Earlier
constructions in \cite{Lok06} use roots of unity of orders $O(n^2)$
but the bounds on rigidity proved there are weaker: $n(n-cr)$ for some
constant $c > 2$.

We pursue a general approach to studying rigidity based on elementary
algebraic geometry and elimination theory. To set up the formalism of
this approach, we begin by reproving Valiant's result that the set of
matrices of rigidity less than $(n-r)^2$ is contained in\footnote{We
  note that this set itself may not be Zariski closed, as was
  mistakenly claimed in some earlier results, e.g., \cite{Lok95},
  \cite{LTV03}. The example in Section~\ref{subsubsec:maxfail} shows
  that the set of matrices of rigidity less than $(n-r)^2$ is not
  Zariski closed.} a proper Zariski closed set in ${\C}^{\nbyn}$,
i.e., such matrices are solutions of a finite system of polynomial
equations. Hence a generic matrix has rigidity at least $(n-r)^2$. In
fact, we prove a more general statement: the set of $n \times n$
matrices of rigidity at most $k$ for target rank $r$ has dimension (as
an affine variety) exactly $n^2 - (n-r)^2 + k$. This sheds light on
the geometric structure of rigid matrices. We believe that our
argument in this context is clearer and cleaner than an earlier work
in the projective setting by \cite{LTV03}. To look for specific
matrices of high rigidity, we consider certain elimination ideals
associated to matrices with rigidity at most $k$. A result in
\cite{DFGS91} using effective Nullstellensatz bounds (for instance, as
in \cite{Bro87,Kol88}) shows that an elimination ideal of a polynomial
ideal must always contain a nonzero polynomial with an explicit degree
upper bound (Theorem~\ref{thm:degree-bound-on-elim-ideals}). We then
use simple facts from algebraic number theory to prove that a matrix
whose entries are primitive roots of unity of sufficiently high orders
cannot satisfy any polynomial with such a degree upper bound. This
gives us the claimed family of matrices of maximum rigidity.

Our primary objects of interest in this paper are the varieties of
matrices with rigidity at most $k$. For a fixed $k$, we have a natural
decomposition of this variety based on the patterns of changes. We
prove that this natural decomposition is indeed a decomposition into
\emph{irreducible} components (Corollary~\ref{coro:decomp}). In fact,
these components are defined by elimination ideals of determinantal
ideals generated by all the $(r+1) \times (r+1)$ minors of an $\nbyn$
matrix of indeterminates. {Better effective upper bounds on the degree
  of a nonzero polynomial in the elimination ideal of determinantal
  ideals than those given by
  Theorem~\ref{thm:degree-bound-on-elim-ideals} would lead to similar
  improvements in the bound on the order of the primitive roots of
  unity we use to construct our rigid matrices.} While determinantal
ideals have been well-studied in mathematical literature, their
elimination theory does not seem to have been {as well-studied}. The
application to rigidity might be a natural motivation for further
investigating the elimination ideals that arise in this situation.

We next consider the question: given a matrix $A$, is there a small
neighbourhood of $A$ within which the rigidity function is
nondecreasing, i.e. such that every matrix in this neighbourhood has
rigidity at least equal to that of $A$? This is related to the notion
of \emph{semicontinuity} of the rigidity function. We give a family of
examples to show that the rigidity function is in general not
semicontinuous. However, the \emph{specific} matrices we produce with
entries being roots of unity as above, by their very construction,
have neighborhoods within which rigidity is nondecreasing.

The rest of the paper is organized as follows. In
Section~\ref{sec:preliminaries}, we introduce some definitions and
notations and recall a basic result from elimination theory. Much of
the necessary background from basic algebraic geometry is reviewed in
Appendix~\ref{sec:prelims}. We introduce our main approach in
Section~\ref{sec:elim}, reprove Valiant's theorem, and compute the
dimension of the variety of matrices of rigidity at most $k$. We
present our new construction of maximally rigid matrices in
Section~\ref{sec:mainresult}. Connection to the elimination ideals of
determinantal ideals is established in Section~\ref{sec:reduction}. In
Section~\ref{sec:topology}, we study semicontinuity of the rigidity
function through examples and counterexamples.

\section{Preliminaries}
\label{sec:preliminaries}
\subsection{Definitions and Notations}
\label{sec:defns-and-notation}

Let \( F \) be a field\footnote{For the most part, we will use the
  field of complex numbers \(\C\). However, many of our definitions
  make sense over an arbitrary field and the theorems we use from
  algebraic geometry hold over any algebraically closed field.}. Then,
by \( M_n(F) \) we denote the algebra of \( n \times n \) matrices
over \( F \). At times, when it is clear from the context, we will
denote \( M_n ( F ) \) by \( M_n \).  We use \( M_{m \times n}(F) \)
to denote the set of \( m \times n \) matrices over \( F \). For \( X
\in M_n (F) \), by \( X_{ij} \) we will denote the \( (i,j) \)-th
entry of \( X \). Given \( X \in M_n (F) \), the support of $X$ is
defined as \( \supp(X) := \{ (i,j) ~| ~ X_{ij} \neq 0 \in F \}.  \)
Given a non-negative integer \( k \), we define
\[
S ( k) := \{ X \in M_n (F)  : | \supp(X)| \leq k \}.
\]
Thus, \( S ( k ) \) is the set of matrices over \( F \) with at most \( k \) non-zero entries.

A \emph{pattern} \( \pi \) is a subset of the positions of an \( n \times n \) matrix.
Then, we define:
\[
S ( \pi ) := \{ X \in M_n (F) : ~ \supp( X ) \subseteq \pi \}.
\]
Note that $S(k) = \bigcup\limits_{|\pi|=k} S(\pi)$.

\begin{definition}
The rigidity function \( \Rig(X,r) \) is the minimum number of entries
we need to change in the matrix \( X \) so that the rank becomes at
most $r$:
\[ \Rig(X,r) := \min \{ \supp(T) \: : \: \rank(X+T) \le r \}. \]
Sometimes, we will allow $T$ to be chosen in $M_n(L)$ for $L$ an
extension field of $F$. In this case we will denote the rigidity by
$\Rig(X,r,L)$.
\end{definition}

Let \( \RIG ( n , r, k) \) denote the set of \( n \times n \) matrices
$X$ such that $\Rig(X,r) = k$. Similarly, we define \( \RIG(n,r,\geq
k) \) to be the set of matrices of rigidity at least $k$ and \(
\RIG(n,r,\leq k) \) to be the set of matrices of rigidity at most $k$.
For a pattern $\pi$ of size $k$, let $\RIG(n,r,\pi)$ be the set of
matrices $X$ such that for some $T_{\pi} \in S(\pi)$ we have
$\rank(X+T_{\pi}) \leq r$. Then we have
\[
\RIG(n,r,\leq k) =   \bigcup_{\pi,|\pi| = k} \RIG(n,r,\pi).
\]

\subsection{Elimination Theory and the Closure Theorem}

We review much of the necessary background from algebraic geometry in
Appendix~\ref{sec:prelims}.  Here we recall a basic result from
Elimination Theory. As the name suggests, Elimination Theory deals
with elimination of a subset of variables from a given set of
polynomial equations and finding the \emph{reduced set} of polynomial
equations (not involving the eliminated variables). The main
results of Elimination Theory, especially the Closure Theorem,
describe a precise relation between the reduced ideal and the given
ideal, and its corresponding geometric interpretation.

Given an ideal \( I = \langle f_1, \ldots, f_s \rangle \subseteq F[
  x_1 , \dots, x_n ] \), the \( l \)-th \emph{elimination ideal} \(
I_l \) is the ideal of \( F[ x_{l+1} , \dots, x_n ] \) defined by
$I_l := I \cap F[ x_{l+1} , \dots, x_n ].$

\begin{theorem} \emph{\bf{(Closure Theorem, page 125, Theorem 3 of \cite{CLO})}}
\label{thm:closure} \\
Let \( I \) be an ideal of \( F[ x_1, \ldots, x_n , y_1, \ldots, y_m ]
\) and \( I_n := I \bigcap F[ y_1, \ldots, y_m ] \) be the \( n \)-th
elimination ideal associated to \( I \). Let \( V(I) \) and \( V(I_n)
\) be the subvarieties of \( \Af^{n+m} \) and \( \Af^m \) (the affine
spaces over $\Fbar$ of dimension \( n+m \) and \( m \) respectively)
defined by $I$ and $I_n$ respectively. Let \( p \) be the natural
projection map from \( \Af^{n+m} \rightarrow \Af^m \) (projection map
onto the \( y \)-coordinates).  Then,
\begin{enumerate}
\item \( V ( I_n ) \) is the smallest (closed) affine variety
  containing \( p ( V(I) ) \subseteq \Af^m \). In other words, \(
  V(I_n ) \) is the Zariski closure of \( p ( V(I)  (\Fbar ) )
  \subseteq \Fbar^m \).
\item When \( V(I)(\Fbar ) \neq \phi \), there is an affine variety \(
  W \) strictly contained in \( V(I_n) \) such that \( V(I_n) - W
  \subseteq p( V(I) ) \).
\end{enumerate}
\end{theorem}

\section{Use of Elimination Theory}
\label{sec:elim}

\subsection{Determinantal Ideals and their Elimination Ideals}
\label{sec:det-ideals}
We would like to investigate the structure of the sets $\RIG(n,r,\leq
k, \Fbar)$ and $\RIG(n,r,\pi, \Fbar)$ and their Zariski closures
\begin{eqnarray*}
\W (n, r, \le k ) & := & \overline{ \RIG(n,r, \le k, \Fbar) } 
\;\;\;
\mbox{ and } \\ \W (n, r, \pi) & := & \overline{ \RIG(n,r, \pi, \Fbar)
}
\end{eqnarray*}
in the $n^2$-dimensional affine space of $n \times n$ matrices. Note
that we have the ``upper bound'' $\RIG(n,r,\leq k) \subset
\RIG(n,r,\leq k, \Fbar)$ and therefore $\overline{\RIG(n,r,\leq k)}
\subset \W (n, r, \le k )$. Let $X$ be an $n\times n$ matrix with
entries being indeterminates $x_1, \ldots, x_{n^2}$. For a pattern
$\pi$ of $k$ positions, let $T_\pi$ be the $n\times n$ matrix with
indeterminates $t_1, \ldots, t_k$ in the positions given by
$\pi$. Note that saying $X + T_{\pi}$ has rank at most $r$ is
equivalent to saying that all its $(r+1) \times (r+1)$ minors
vanish. Let us consider the ideal generated by these minors:
\begin{equation}
\label{eq:Inrpi}
  I(n,r,\pi)  :=  \left\langle \minors_{(r+1)\times(r+1)}(X + T_\pi ) \right\rangle
  \subseteq  F[x_1, \ldots , x_{n^2}, t_1 , \ldots , t_k].
\end{equation}
It then follows from the definition of rigidity that $\RIG(n,r,\pi, \Fbar)$
is the projection from $\Af^{n^2} \times \Af^k$ to $\Af^{n^2}$ of the
algebraic set $V(I(n,r,\pi))(\Fbar)$. Thus, if we define the elimination
ideal
\[
  EI(n,r,\pi) := I(n,r,\pi) \cap F[x_1 , \ldots , x_{n^2}] \subseteq
  F[x_1, \ldots , x_{n^2}],
\]
then by the Closure Theorem (Theorem~\ref{thm:closure}), we obtain
\begin{equation}
\label{eq:W-variety}
\W(n,r,\pi) =  V(EI(n,r,\pi)).
\end{equation}
Note that
\[
\W(n,r,\le k) =  \bigcup_{\pi,|\pi| = k} \W(n,r,\pi).
\]

\subsection{Valiant's Theorem}
\label{sec:valiant-thm}

The following theorem due to Valiant \cite[Theorem 6.4, page
  172]{Val77} says that a generic matrix has rigidity $(n-r)^2$. That
is, for $k < (n-r)^2$, the dimension of \( \W(n,r,\leq k) \) is
strictly less than \( n^2 \).

A reader familiar with Valiant's proof will realize that our
proof is basically a rephrasing of Valiant's proof in the language of
algebraic geometry. The point of this proof is to set up the formalism
and use it later; in particular, when we compute the exact dimension
of the rigidity variety \( \W(n,r,\leq k) \).

\begin{theorem}\emph{\bf (Valiant)}
\label{thm:dim-upper-bound-valiant}
Let \( n \geq 1, 0 < r < n \) and \( 0 \le k < (n-r)^2 \). Let \( \W
:= \W (n, r, \le k ) \) be as above. Then,
\[
\dim( \mathcal{W} ) < n^2 .
\]
\end{theorem}
\begin{proof}
  Let \( \pi \subseteq \{(i,j) \,| \, 1\leq i,j \leq n\}\) be a
  pattern of size \( k \).  For a choice of $0 \leq s \leq r$, we let
  \( \tau \) denote a choice of $s$ rows and $s$ columns, and for a
  matrix $B$, let $B_{\tau}$ be the corresponding submatrix of $B$,
  whose determinant is one of the $s \times s$ minors of $B$. For $s =
  0$, we let $B_\tau$ be the empty matrix, with determinant defined to
  be $1$.

  For $s \leq r$, define \( \RIG ( n, s, \pi, \tau ) \) to be the set
  of all \( n \times n \) matrices \( A \) that satisfy the following
  properties: there exists some \( n \times n \) matrix \( T_\pi \)
  such that

\begin{enumerate}
\item \( \supp(T_\pi ) \subseteq \pi \),
\item \( \rank (A + T_\pi ) = s \), and
\item \( \det ( ( A + T_\pi )_\tau) \neq 0 \) where \( \tau \) denotes
  the fixed \( s \times s \) minor as above.
\end{enumerate}

Recall that \( S (\pi) \) is the set of matrices whose support is
contained in \( \pi \). Let us also define
\[
\RANK(n,s,\tau) := \{ C \in M_n ~ | ~ \rank(C) = s ~ \textrm{and}
\det(C_\tau) \neq 0 \}.
\]
By definition, every element \( A \in \RIG(n,s, \pi,\tau) \) can be
written as \( C - T_\pi \), with \( C \in \RANK(n, s , \tau ) \) and
\( T_\pi \in S(\pi) \).

We first prove the following lemma:

\begin{lemma}
\label{lem:rank-variety-dimension}
\(
	\dim( \RANK(n,s,\tau) ) = n^2 - (n-s)^2.
\)
\end{lemma}

\begin{proof}
Without loss of generality we can assume that \( \tau \) is the upper
left \( s \times s \)-minor. Thus we can write a \( C \in
\RANK(n,s,\tau) \) as
\[
C = \left[
\begin{matrix}
C_{11} & C_{12} \\
C_{21} & C_{22}
\end{matrix}
\right],
\]
where \( \rank(C) = s \) and \( C_{11} \) is an \( s \times s \)
matrix whose determinant is non-zero.

Since the matrix $C_{11}$ is nonsingular of dimension equal to $s =
\rank(C)$, it follows that the first $s$ columns are linearly
independent and span the column space of $C$. Therefore each of the
last $(n-s)$ columns is a linear combination of the first $s$ columns
in exactly one way, and the linear combination is determined by the
entries of $C_{12}$. Formally, we have the equation
$$
 C_{22} = C_{21} C_{11}^{-1} C_{12}.
$$ 
The set of all $C_{11}$ is an affine open set of dimension $s^2$
 and $C_{12}$ and $C_{21}$ can each range over
 ${\Af}^{s(n-s)}$. Hence, the algebraic set $\RANK(n,s,\tau)$ has
 dimension exactly $s^2+2r(n-s) = n^2 - (n-s)^2$.
\end{proof}

Consider the following natural map $\Phi$:
\begin{equation}
\label{eq:valiant-map}
  \Af^{n^2-(n-s)^2} \times \Af^{k} \supset \RANK(n,s,\tau) \times
  S(\pi) \stackrel{\Phi}{\longrightarrow} M_n \cong \Af^{n^2},
\end{equation}
taking $(X, T_\pi)$ to $X+T_\pi$. The image of \( \Phi \) is exactly
\( \RIG(n,r,\pi,\tau) \) as defined at the beginning of this proof.

Also, note that \( \dim ( S(\pi) ) = |\pi| .  \) We note that if there
is a surjective morphism from an affine variety $X$ to another affine
variety $Y$, then $\dim Y \leq \dim X$ (a more formal statement
appears as Lemma~\ref{lem:dim-image-rational-isomorphism} in
Appendix~\ref{sec:prelims}).  Thus for \( k \le (n-s)^2-1 \), we get
\begin{equation}\label{eqn:dimupperbound}
\dim( \overline{\image(\Phi)} ) = \dim ( \overline{\RIG(n,s,\pi,\tau)}
) \leq n^2-(n-s)^2+k < n^2 .
\end{equation}
Note that
\begin{equation}\label{eqn:W}
\W = \bigcup_{s \leq r, \tau, \pi} \overline{\RIG(n,s,\pi,\tau)}
\end{equation}
and that completes the proof of the theorem.
\end{proof}

Thus we have proved that the set of matrices of rigidity strictly
smaller than \( (n-r)^2 \) is contained in a proper closed affine
variety of \( \Af^{n^2} \), and thus is of dimension strictly less
than \( n^2 \). In other words, a {\em generic matrix}, i.e. a matrix
that lies outside a certain proper closed affine subvariety of \(
\Af^{n^2}\), is \emph{maximally rigid} (even if we allow changes by
elements of $\Fbar$, rather than just $F$). Therefore, over an
infinite field $F$ (for instance, an algebraically closed field),
there always exist maximally rigid matrices.

We now refine Valiant's argument and prove the following exact bound
on the dimension of $\W$. The main point of the proof is a \emph{lower
  bound} on $\dim (\W)$.
\begin{theorem}
\label{thm:dim-of-rigid-matrices}
Let \( 0 \le r \le n \) and \( 0 \leq k \leq ( n-r)^2 \). Then
\[
 	\dim ( \W ) = n^2 - ( n -r)^2 + k.
\]
\end{theorem}

\begin{proof}

By the above proof of Theorem \ref{thm:dim-upper-bound-valiant} (see
Equation (\ref{eqn:dimupperbound})), we only need to prove that the \(
\dim ( \cal{W} ) \) is at least \( n^2 - ( n - r)^2 + k \). By
Equation (\ref{eqn:W}) as above,
\[
\dim ( \W ) = \max_{s \leq r, \pi, \tau } \dim ( \overline{\RIG(n,s,\pi,\tau)} ).
\]
Thus, to prove the theorem it is sufficient to prove that
for some $r \leq s$, and some \( \pi \) and \( \tau \):
\[
\dim ( \RIG(n,s,\pi,\tau ) ) \geq n^2 - ( n - r)^2 + k.
\]
We take $s = r$ and choose \( \pi \) and \( \tau \) as follows. Fix a
pattern \( \pi \) of size \( k \) such that it is a subset of \( \{
(i,j) \, | \, r + 1 \leq i , j \leq n \} \). This is possible because
\( k \leq ( n - r)^2 \). Let \( \tau \) be the top left \( r \times r
\) minor. We now define:
\begin{equation}
U := \left\{
\left[\begin{matrix}
G & A \\
B & X_\pi + B G^{-1} A
\end{matrix} \right]
~:~ G \in GL_r , ~ A \in M_{r \times (n-r)} , ~ B \in M_{(n-r) \times r }, ~ X_\pi \in S( \pi )
\right\}.
\end{equation}
As an affine algebraic variety, \( U \) is
isomorphic to \( GL(r) \times \Af^{n \times (n-r) } \times \Af^{(n-r) \times r } \times \Af^k \), and
thus \( \dim (U ) = r^2 + 2 ( n-r) r + k = n^2 - (n-r)^2 + k \).
If we subtract the matrix
$$
\left[ \begin{matrix}
0 & 0 \\
0 & X_\pi
\end{matrix} \right]
$$
from the matrix above, we get a matrix
$$
\left[ \begin{matrix}
G & A \\
B & B G^{-1} A
\end{matrix} \right]
$$ of rank exactly $r$ since the the first $r$ columns are linearly
independent ($G$ being invertible) and the last $n-r$ columns are a
linear combination of the first $r$, obtained by multiplying on the
right by the matrix $G^{-1} A$. Therefore, \( U \subseteq
\RIG(n,r,\pi,\tau ) \), and hence \( \dim ( \RIG(n,r,\pi,\tau ) ) \geq
n^2 - (n-r)^2 + k.  \)
\end{proof}

\begin{remark}
\emph{A similar argument or line of study - though in the projective
  setting - is also found in \cite{LTV03}. Our formalism and proofs
  seem clearer and simpler. Our theorem is also very explicit.}
\end{remark}

\subsection{Rigid Matrices over the field of Complex Numbers}
\label{sec:mainresult}

Recall that to say that the rigidity of a matrix \( A \) for target
rank \( r \) is at least \( k \), it suffices to prove that the matrix
\( A \) is not in \( \W (n, r, \le (k-1)) \). We use this idea to
achieve the maximum possible lower bound for the rigidity of a family
of matrices over the field of complex numbers \( \C \). As a matter of
fact, we obtain matrices with real algebraic entries with rigidity
$(n-r)^2$.
\begin{theorem}
\label{thm:rigid-matrices-over-C}
Let \( \Delta (n) = n^{4n^2} \) and let \( p_{i,j} > \Delta (n) \) be
distinct primes for $1 \le i,j \le n$.  Let \( K = \Q(\zeta_{1,1} ,
\ldots , \zeta_{n,n} ) \) where \( \zeta_{i,j} = e^{ {2\pi
    \sf{i}}/{p_{i,j}}} \).  Let \( A (n) := [ \zeta_{i,j} ] \in M(n,K)
\). Then, for any field \( L \) containing \( K \),
\[
  \Rig ( A(n),r, L ) = (n-r)^2 .
\]
\end{theorem}
\begin{proof}
For simplicity, we will index the \( \zeta_{i,j} \) by \( \zeta_\alpha
\) for \( \alpha = 1 \) to \( n^2 \), and similarly \( p_\alpha
\). First, note that we may assume $1 \leq r \leq n-1$ since for $r =
n$ the statement of the theorem is a tautology, and for $r = 0$, it is
obvious.  We prove the theorem by showing that
\[
  A ( n ) \notin \W (n, r , \leq (n-r)^2-1  ) (L) .
\]
Thus it is sufficient to prove that
\[
A (n) \notin \W (n, r, \pi )(L)
\]
for any pattern \( \pi \) with \( |\pi| = k := \kM - 1 \). Let $\pi$
be any such pattern. To simplify notation, let us define $\W := \W (n,
r, \pi )(L)$.  By Theorem~\ref{thm:dim-upper-bound-valiant} we have:
\[
\dim ( \W ) \leq \dim( \W( n, r, \leq (n-r)^2-1) ) < n^2 .
\]
Equivalently (by Hilbert's Nullstellensatz),
\[
EI (n, r, \pi ) \ne (0) .
\]
Proving that \( A (n) \notin \W \) is equivalent to showing the
existence of a \( g \in EI ( n, r, \pi ) \) such that \( g ( A(n) )
\ne 0 \). The key to the proof of the theorem is to produce a
polynomial $g$ of sufficiently low degree.

\begin{claim}
There is a polynomial $g \in EI(n,r,\pi)$ of total degree less than
$\Delta(n)$.
\end{claim}

To prove the claim, we use the following theorem:
\begin{theorem}\emph{\bf (\cite{DFGS91}, Proposition 1.7 and Remark 1.8)}
\label{thm:degree-bound-on-elim-ideals}
Let \( I = \langle f_1, \ldots , f_s \rangle \) be an ideal in the
polynomial ring \( F [ Y ] \) over an infinite field \( F \), where \(
Y = \{ y_1, \ldots , y_m \} \). Let \( d_{\max} \) be the maximum
\emph{total degree} of a generator \( f_i \). Let \( Z = \{ y_{i_1},
\ldots , y_{i_\ell } \} \subseteq Y \) be a subset of indeterminates
of \( Y \). If \( I \cap F [ Z ] \ne (0) \) then there exists a
non-zero polynomial \( g \in I \cap F [Z] \) such that, \( g =
\sum_{i=1}^{s} g_if_i, \) with \( g_i \in F [Y] \) and \( \deg (g_i
f_i) \le d^m ( d^m + 1), \) where $d = \max ( d_{\max}, 3)$.
\end{theorem}
\begin{remark}
\emph{ Note that the proof of Theorem
  \ref{thm:degree-bound-on-elim-ideals} relies on a slightly different
  notion of the degree of a variety than the usual definition in
  projective algebraic geometry. This definition was used in
  \cite{Hei83} to prove the {\em B\'ezout inequality}. For an
  explanation of how the first sentence of Remark 1.8 of \cite{DFGS91}
  follows from this inequality, we refer the reader to Proposition 2.3
  of \cite{HS82}.  }
\end{remark}

Let us apply Theorem~\ref{thm:degree-bound-on-elim-ideals} to our case
- in the notation of this theorem our data is as follows: \( F := \Q
\), \( Y := \{ x_1, \ldots, x_{n^2} , t_1 , \ldots, t_k \} \), \( Z :=
\{ x_1, \ldots, x_{n^2} \} \), \( \Sigma_{r+1} := \) set of all minors
of size \( (r+1) \), \( f_\tau := \det ( ( X + T_\pi )_{\tau} ) \) for
\( \tau \in \Sigma_{r+1} \), where by \( Y_\tau \) we denote the \(
\tau \)-th minor of \( Y \), and \( I := I(n, r, \pi ) = \langle
f_\tau : ~\tau \in \Sigma_{r+1} \rangle \) as defined in
(\ref{eq:Inrpi}).
We may as well assume $n \geq 3$, since for $n = 2$ the claim is easy
to verify by explicit calculation. Then we have:
\begin{eqnarray*}
m &=& n^2 + \kM - 1 \leq 2n^2 - 2, \\
d &=& \max(r+1,3) \le n,  \;\; \mbox{ and} \\
I \cap F [ Z ] & = & EI (n,r,\pi ) \ne (0).
\end{eqnarray*}
By Theorem~\ref{thm:degree-bound-on-elim-ideals} there exists a
\[
g \neq 0 \in EI (n,r,\pi )  \subseteq \Q[x_1 , \ldots , x_{n^2} ]
\]
such that
\[
\deg (g) \leq d^m (d^m + 1) \leq n^{2n^2 - 2} (n^{2n^2 - 2} + 1) <
n^{4n^2} = \Delta (n) .
\]
We will now apply the following
Lemma~\ref{lem:nonzero-poly-at-alg-nums}, which we prove later, to
this situation.
\begin{lemma}
\label{lem:nonzero-poly-at-alg-nums}
Let \( N \) be a positive integer. Let \( \theta_1 , \cdots, \theta_m
\) be \( m \) algebraic numbers such that for any \( 1 \le i \le m \),
the field \( \Q ( \theta_i ) \) is Galois over \( \Q \) and such that
\[
	[ \Q ( \theta_i ) : \Q ] \ge N \;\; \mbox{ and } \;\;
	\Q ( \theta_i ) \cap \Q ( \theta_1, \ldots, \theta_{i-1} ,
 	\theta_{i+1} , \ldots, \theta_m ) = \Q .
\]
Let \( g ( \underline{x} ) \neq 0 \in \Q [ x_1 , \ldots , x_m ] \) such that
\( \deg ( g) < N \). Then,
\(
g( \theta_1 , \ldots, \theta_m ) \neq 0 .
\)
\qed
\end{lemma}
Let us set \( m =n^2, N = \Delta(n), l := \deg(g) \le N \) in
Lemma~\ref{lem:nonzero-poly-at-alg-nums}. It is now easy to check that
\[
[\Q (\zeta_\alpha ) : \Q ] = p_\alpha - 1 \geq \Delta(n) = N
\]
and
\[
\Q ( \zeta_\alpha ) \cap \Q ( \zeta_1, \ldots, \zeta_{\alpha -1} ,
 	\zeta_{\alpha + 1} , \ldots, \zeta_{n^2} ) = \Q .
\]
The latter follows from the fact that the prime \( p_\alpha \) is
totally ramified in \( \Q ( \zeta_\alpha ) \) and is unramified in \(
\Q ( \zeta_1, \ldots, \zeta_{\alpha -1} , \zeta_{\alpha + 1} , \ldots,
\zeta_{n^2} ) \); see Theorem 4.10 in \cite{Nark}. Thus
Lemma~\ref{lem:nonzero-poly-at-alg-nums} is applicable and we get:
\[
		g ( \zeta_1, \ldots , \zeta_{n^2} ) \neq 0.
\]

To complete the argument (for Theorem
~\ref{thm:rigid-matrices-over-C}), now we prove
Lemma~\ref{lem:nonzero-poly-at-alg-nums}.

{\bf Proof of Lemma~\ref{lem:nonzero-poly-at-alg-nums}}: 
By induction on $m$. For \( m = 1 \) this is trivial.  Now suppose
that the statement is true when the number of variables is strictly
less than \( m \).  Assuming that the statement is not true for \( m
\), we will arrive at a contradiction. This will prove the lemma.

Let \( g \in \Q [ \underline{x} ] \) with \( l := \deg (g ) < N \) be
such that
\[
g( \theta_1 , \ldots, \theta_m ) = 0 ,
\]
with \( \theta_i \), \( 1 \leq i \leq m \), satisfying the conditions
as in the theorem. Since the statement is true for \( (m-1) \)
variables by the inductive hypothesis, without loss of generality, we
can assume that all the variables and hence \( x_m \) appears in \( g
\). Let us denote \( x_m \) by \( x \). Let us write
\[
 g(x_1, \ldots, x_m ) = \sum_{i=0}^l f_i (x_1 , \ldots, x_{m-1} ) x^{l
   - i} .
\]
Note that \( l < N \) and \( \deg (f_i ) < N \) for \( 0 \le i \le l
\). Since \( g \neq 0 \), for some \( i , ~0 \le i \le l \) the
polynomial \( f_i \neq 0 \). Thus, by the inductive hypothesis,
\[
f_i ( \theta_1, \ldots , \theta_{m-1} ) \neq 0 .
\]
Thus \( g ( \theta_1, \ldots, \theta_{m-1} )(x) \neq 0 \in \Q (
\theta_1, \ldots, \theta_{m-1} )[x] \).  This implies that \( \theta_m
\) satisfies a non-zero polynomial over \( \Q ( \theta_1, \ldots,
\theta_{m-1} ) \) of degree \( \leq l < N \). Thus:
\begin{equation}
\label{eq:field-ext-bound}
[ \Q ( \theta_1, \ldots, \theta_m ) : \Q ( \theta_1, \ldots,
  \theta_{m-1} ) ] \le l < N .
\end{equation}
On the other hand, since \( \Q ( \theta_m ) \cap \Q ( \theta_1,
\ldots, \theta_{m-1} ) = \Q \) and the fields \( \Q ( \theta_i ) \)
are Galois over \( \Q \), by Theorem~\ref{thm:Galois-parallelogram}
(stated below), we conclude that
\[
[ \Q ( \theta_1, \ldots, \theta_{m-1} ) ( \theta_m ) : \Q ( \theta_1,
  \ldots, \theta_{m-1}) ] = [ \Q ( \theta_m ) : \Q ] \ge N.
\]
This contradicts (\ref{eq:field-ext-bound}) above and proves the lemma.

\begin{theorem}\emph{\bf (\cite{Lang}, Theorem 1.12, page 266)}
\label{thm:Galois-parallelogram}
Let \( K \) be a Galois extension of \( k \), let \( F \) be an
arbitrary extension of $k$, and assume that \( K \), \( F \) are
subfields of some other field. Then \( KF \) (the compositum of \( K
\) and \( F \)) is Galois over \( F \), and \( K \) is Galois over \(
K \bigcap F \). Let \( H \) be the Galois group of \( KF \) over \( F
\), and \( G \) the Galois group of \( K \) over \( k \). If \( \sigma
\in H \) then the restriction of \( \sigma \) to \( K \) is in \( G
\), and the map $\sigma \mapsto \sigma |_K$ gives an isomorphism of \(
H \) on the Galois group of \( K \) over \( K \cap F \). In
particular, $[KF : F ] = [ K : K \cap F ]$.
\end{theorem}
This concludes the proof of Theorem~\ref{thm:rigid-matrices-over-C}.
\end{proof}

Note that Theorem~\ref{thm:rigid-matrices-over-C} is true for any
family of matrices \( A(n) = [ \theta_{i,j} ] \) provided the \(
\theta_{i,j} \) satisfy Lemma~\ref{lem:nonzero-poly-at-alg-nums}.
Hence, we have:
\begin{corollary}
Let \( A(n) := [\zeta_{i,j} + \overline{\zeta_{i,j}} ] \), where \(
\zeta_{i,j} \) are primitive roots of unity of order \( p_{i,j} \)
such that \( p_{i,j} - 1 \ge 2 \Delta(n) \) (here \(
\overline{\zeta_{i,j}} \) denotes the complex conjugate of \(
\zeta_{i,j} \)). Then, \( A(n) \in M(n, \R) \) has \( \Rig( A(n) , r )
= (n- r)^2 \).
\end{corollary}

\begin{proof}
We apply the remark above with $\theta_{i,j} = \zeta_{i,j} +
\overline{\zeta_{i,j}}$, which generates the maximal real subfield of
$\Q(\zeta_{i,j})$. These fields are Galois over $\Q$, and since
$\Q(\theta_{i,j}) \subset \Q(\zeta_{i,j})$, they satisfy the linear
disjointness property which forms the second part of the assumption of
Lemma 10.
\end{proof}

\section{Reduction to Determinantal Ideals}
\label{sec:reduction}

In this section, we show that the natural decomposition of the
rigidity varieties $\W(n,r, \leq k) = \bigcup_{|\pi|=k} \W(n,r,\pi)$ is
indeed a decomposition into \emph{irreducible} affine algebraic varieties. In
fact, these components turn out to be varieties defined by elimination
ideals of determinantal ideals generated by all the $(r+1) \times
(r+1)$ minors.

To improve the bounds on the orders of primitive roots of unity in
Theorem~\ref{thm:rigid-matrices-over-C}, it suffices to improve the
degree bounds given by Theorem~\ref{thm:degree-bound-on-elim-ideals}
for the special case when $I$ is a determinantal ideal. However, we do
not know of such an improvement even for the special case when $I$ is
the determinantal ideal of a generic Vandermonde matrix.

To show the decomposition, we will continue to use the notation from
Section~\ref{sec:elim}. Consider the matrix \( X + T_\pi \). Let \( x
= \{x_1,\ldots,x_{n^2} \} = x_{\bar{\pi}} \bigcup x_\pi \), where \(
x_\pi \) is the set of variables that are indexed by \( \pi \) and \(
x_{\bar{\pi }}\) is the set of remaining variables.

Let
\[
J := I(n, r, \pi) = \left\langle \minors_{(r+1)\times(r+1)} (X + T_\pi ) \right\rangle
\]
be the ideal of $\Q[x,t] = \Q[x_{\pi}, x_{\bar{\pi}}, t_{\pi}]$
generated by the $(r+1) \times (r+1)$ minors of $X + T_{\pi}$. Let
\[
\begin{array}{rcl}
        J_1     & := & J \cap \Q[x_\pi, x_{\bar{\pi}} ] \subseteq  \Q [ x_1, \ldots , x_{n^2} ],  \\
        J_2       & := & J_1 \cap \Q[ x_{\bar{\pi}} ], \\
     I_{r+1}  & := & \left\langle \minors_{(r+1)\times(r+1)}(X) \right\rangle \subseteq \Q [ x ], \;\;\; \mbox{ and } \\
  EI_{r+1}  & := & I_{r+1} \cap \Q[x_{\bar{\pi}}] \subseteq \Q [ x_{\bar{\pi}} ].
\end{array}
\]
Notice that since $J_1$ is the elimination ideal of $J$
w.r.t. eliminating variables $t_{\pi}$, a matrix $A$ lies in $\W (n,
r, \pi ) = \overline{ \RIG(n,r, \pi, \Fbar)}$ if and only if its
entries lie in the variety defined by the ideal $J_1$. Therefore,
$J_1$ equals the elimination ideal $EI(n,r,\pi)$ defined in Section
\ref{sec:det-ideals}, by definition. Also, $I_{r+1}$ is the ideal
generated by the $(r+1) \times (r+1)$ minors of $X$ and $EI_{r+1}$ its
elimination ideal for the polynomial ring over the rationals generated
by the variables $x_{\bar{\pi}}$.

\begin{proposition}
\label{prop:detred}
$J_1 = J_2\Q[x]$ (the ideal generated by $J_2$ in $\Q[x]$) and $J_2 = EI_{r+1}$.  In particular, $EI(n,r,\pi) = EI_{r+1} \Q[x]$ considered as ideals in $\Q[x]$.
\end{proposition}
\begin{proof}

  First, notice that in the $(r+1) \times (r+1)$ minors of $X +T_\pi$,
  the variable $t_{i,j}$, for $(i,j) \in \pi$, always occurs in
  combination with $x_{i,j}$ as $t_{i,j} + x_{i,j}$. Therefore,
  eliminating the variables $t_\pi$ will also automatically eliminate
  the variables $x_\pi$, giving the equality of the generators of the
  ideals $J_1$ and $J_2$. Therefore $J_1 = J_2 \Q[x]$. More formally,
  consider the automorphism $\phi$ of $\Q[x_\pi, x_{\bar{\pi}},
  t_\pi]$ defined by letting $\phi(t_{i,j}) = x_{i,j} + t_{i,j}$ for
  each $(i,j) \in \pi$ and $\phi(x_{i,j}) = x_{i,j}$ for all
  $(i,j)$. The ideal $J_1 = J \cap \Q[x_\pi, x_{\bar{\pi}} ] \subseteq
  \Q [ x_1, \ldots , x_{n^2} ]$ must equal the ideal $\phi(
  \phi^{-1}(J) \cap \phi^{-1}\Q[x_1, \ldots, x_{n^2}])$, since $\phi$
  is an isomorphism. But $\phi^{-1}(J)$ is generated by determinants
  of matrices only involving the variables $t_\pi$ and
  $x_{\bar{\pi}}$, whereas $\phi^{-1}\Q[x_1, \ldots, x_{n^2}]) =
  \Q[x_1, \ldots, x_{n^2}]$, so that $\phi^{-1}(J) \cap
  \phi^{-1}\Q[x_1, \ldots, x_{n^2}]$ is generated by polynomials only
  involving the variables of $x_{\bar{\pi}}$. Therefore
  $\phi^{-1}(J_1) = \phi^{-1}(J) \cap \phi^{-1}\Q[x_1, \ldots,
  x_{n^2}] = J_2 \Q[x]$. Taking the image under $\phi$, we get $J_1
  = J_2 \Q[x]$.

  The equation $J_2 = EI_{r+1}$ follows from similar considerations,
  noting that the variables $x_{i,j}$ for $(i,j) \in \pi$ always occur
  in the combination $x_{i,j} + t_{i,j}$ in the minors which generate
  $J$. Therefore eliminating them eliminates $t_{i,j}$ as well. More
  formally, consider the isomorphism $\psi:\Q[x_\pi, x_{\bar{\pi}},
  t_\pi] \rightarrow \Q[x_{\pi}, x_{\bar{\pi}}, t_\pi]$ defined by
  letting $\psi(x_{i,j}) = x_{i,j} + t_{i,j}$ for each $(i,j) \in
  \pi$, while $\psi(t_{i,j}) = t_{i,j}$ for $(i,j) \in \pi$ and
  $\psi(x_{i,j})= x_{i,j}$ for $(i,j) \not \in \pi$.  Then again we
  have $J_2 = J_1 \cap \Q[x_{\bar{\pi}}] = J \cap \Q[x_{\bar{\pi}}] =
  \psi(\psi^{-1}(J) \cap \psi^{-1}(\Q[x_{\bar{\pi}}])) = \phi(I_{r+1}
  \Q[x,t_\pi] \cap \Q[x_{\bar{\pi}}]) = \phi(EI_{r+1}) = EI_{r+1}
  \subset \Q[x_{\bar{\pi}}]$.
\end{proof}

The following is a well-known theorem; see \cite[Theorem 1]{HE71} and
\cite[Chapter 2]{BV}.
\begin{theorem}
\label{thm:det-irred}
Let \( \RANK(n, \leq r) \) be the set of all rank \( \leq r \)
matrices of \( M_n \cong \Af^{n^2} \).  Then
\begin{enumerate}
\item \( I(\RANK(n, \leq r)) = I_{r+1} \) and \( \RANK(n, \leq r) = V(I_{r+1})\).
\item \( I_{r+1} \) is a prime ideal of \( \Q [ X ] \). In particular,
  \( \RANK(n, \leq r) \) is an irreducible variety.
\end{enumerate}
\end{theorem}
\begin{corollary}
\label{coro:decomp}
In the natural decomposition $\W(n, r, \le k) = \bigcup\limits_{|\pi|=k}
\W(n,r,\pi)$, the $\W(n,r,\pi)$ are irreducible varieties.
\end{corollary}
\begin{proof}
In general if $J$ is a prime ideal of a commutative ring $S$ and if
$R$ is a subring of $S$, then $I = J \cap R$ is prime ideal of
$R$. Using this, it follows that the elimination ideal $EI_{r+1}
\subseteq \Q[x_{\bar{\pi}}]$ is a prime ideal since $I_{r+1} \subseteq
\Q[x]$ is a prime ideal by Theorem~\ref{thm:det-irred}.

By Lemma~\ref{prop:detred}, $EI(n,r,\pi) = EI_{r+1}\Q[x]$ considered
as ideals in $\Q[x]$. We need to prove that $EI(n,r,\pi)$ is a prime
ideal in $\Q[x]$. To prove this we use the following general fact: if
$S = R[y]$ where $y$ is transcendental over an integral domain $R$
then, $IS$, the ideal generated by $I$ in $S$, is a prime ideal of
$S$. To see this, note that $S/IS \cong (R/I)[y]$. Now, $R/I$ is an
integral domain (this is equivalent to $I$ being prime), therefore so
is $(R/I)[y]$. Therefore $IS$ is a prime ideal. Now let $R =
\Q[x_{\bar{\pi}}]$ and $S = \Q[x] = R[x_{\pi}]$. Let $I = EI_{r+1}$
which is a prime ideal of $R$. Then, $IS = EI_{r+1}\Q[x] =
EI(n,r,\pi)$ (Lemma~\ref{prop:detred}) and further more, from the
general comments as above, it follows that the latter is a prime ideal
in $\Q[x]$. Thus, $W(n,r,\pi) = V(EI(n,r,\pi)) = V(EI_{r+1})$ (by
(\ref{eq:W-variety})) is an irreducible subvariety of $\Af^{n^2}$.
\end{proof}

Finally, we end with the observation that Proposition
\ref{prop:detred} gives us a slight improvement on Theorem
\ref{thm:rigid-matrices-over-C}.

\begin{theorem} \label{mainthm}
Let \( \Delta (n) = 2n^{2n^2}. \) Let \( p_{i,j} \) for $1 \le i,~j
\le n$ be distinct primes such that \( p_{i,j} > \Delta (n) \).  Let
\( K = \Q(\zeta_{1,1} , \ldots , \zeta_{n,n} ) \) where \( \zeta_{i,j}
= e^{ {2\pi \sf{i}}/{p_{i,j}}} \).  Let \( A (n) := [ \zeta_{i,j} ] \in
M(n,K) \). Then, for any field \( L \) containing \( K \),
\[
  \Rig ( A(n),r, L ) = (n-r)^2 .
\]
\end{theorem}

\begin{proof}
  The only change is the improvement on $\Delta(n)$, which follows
  from Theorem \ref{thm:degree-bound-on-elim-ideals} as before, using
  the fact that $EI(n,r,\pi) = EI_{r+1} \Q[x]$ by Proposition
  \ref{prop:detred} above. Since now there are only $m = n^2$
  variables in all, we easily get the bound $\deg(g) \leq n^{n^2}
  (n^{n^2} + 1) < \Delta(n)$. (As before, we have assumed $n \geq 3$.)
\end{proof}

\section{Topology of Rigidity with some Examples}
\label{sec:topology}
In this section, we make some observations about the topological
behavior of the rigidity function in $M_n(\C)$. The main motivation is
to examine if all matrices within a small neighborhood of a matrix $A$
are at least as rigid as $A$. For instance, the matrices $A(n)$ from
Theorem~\ref{thm:rigid-matrices-over-C} have an open neighborhood
around them within which the rigidity function is constant. This is a
direct consequence of their very construction since they are outside
the closed sets $\W(n,r,\leq (n-r)^2-1)$. We ask if this is a general
property of the rigidity function itself. The notion of
\emph{semicontinuity} of a function captures this property.

\subsection{Semicontinuity of Rigidity}
\label{sec:semi}
Intuitively, if a function is (lower) semicontinuous at a given point,
then within a small neighborhood of that point, the function is
nondecreasing. Formally,
\begin{definition}
\textbf{Semicontinuity:} Let $Y$ be a topological space. A function
$\phi: Y \rightarrow \Z$ is \emph{(lower) semicontinuous} if, for each
$n$, the set $\{y \in Y : \phi(y) \leq n\}$ is a closed subset of
$Y$. That is, for each $y$ there is a neighbourhood $U$ of $y$ such
that for $y' \in U, \phi(y') \geq \phi(y)$.
\end{definition}
The rank function of a matrix, for example, is a lower semicontinuous
function on the space of all $\nbyn$ complex matrices. Unfortunately,
the rigidity function does not in general have this nice property. We
now show below that that there is an infinite family of matrices
$\{A_n \}_{n \ge 1}$ such that, for all $n$ and any $\epsilon_n > 0$,
there is a matrix $B_n$ that is $\epsilon_n$-close to $A_n$ but having
rigidity strictly \emph{smaller} than that of $A_n$.

We start with a $3 \times 3$ example. Let \( a, b, c, d, e \) be
non-zero rational numbers and consider
\begin{equation}
\label{eq:matrix}
A = \left[
\begin{matrix}
a & b & c \\
d & 0 & 0 \\
e & 0 & 0
\end{matrix}
\right] \in M_3(\C).
\end{equation}
Observe that \( \rk(A) = 2 \) and by changing two (and no fewer)
entries its rank can be brought down to 1. Hence, \( \Rig(A,1) = 2 \).

Now for any \( \epsilon > 0 \), let
\[
A(\delta) =
\left[
\begin{matrix}
a & b & c \\
d & bd\delta & cd\delta \\
e & be\delta & ce\delta
\end{matrix}
\right],
\]
where \( \delta \neq 0 \) and \( \delta \neq 1/a \), be such that \(
\epsilon \ge \max \{bd\delta, cd\delta, be\delta, c e \delta \}
\). Note that \( \rk(A(\delta)) = 2 \). Also \( \Rig (A(\delta),1) = 1
\) because changing \( a \) to \( \frac{1}{\delta} \) will make all
the \( 2 \times 2 \) sub-determinants of \( A(\delta) \) zero. Thus,
we have a matrix \( A(\delta) \) which is in the open \( \epsilon
\)-ball around \( A \) such that \( \Rig(A,1) > \Rig(A(\delta),1)
\). This proves conditions for semicontinuity of rigidity do
\emph{not} hold at $A$.

To produce an infinite family for any given $n$, take $\alpha, a_1,
b_1, \dots, a_{n-1}, b_{n-1}$ to be non-zero rational numbers, and let
\[
A_n :=
\left[
	\begin{matrix}
				\alpha & a_1 & a_2 & & \ldots & & a_{n-1} \\
					 b_1 & 0 & 0 & & \ldots & & 0 \\
					 b_2 & 0 & 0 & & \ldots & & 0 \\
		         . & . & . & & \ldots & & . \\
		         . & . & . & & \ldots & & . \\
	         b_{n-1} & 0 & 0 & & \ldots & & 0
	\end{matrix}
\right] \in M_n(\C).
\]
Then, it is easy to show by induction that
for \( n \ge 3 \), \( \rk(A_n) = 2 \), and \( \Rig(A_n,1) = n-1 \).

On the other hand, for a given \( \epsilon \), choose a \( \delta \)
such that \( \epsilon \ge \max_{i,j} \{a_ib_j\delta\} \) with \(
\delta \neq 0, ~ 1/\alpha \) and let
\[
A_n (\delta)  = \left[
\begin{matrix}
 \alpha & a_1 & a_2 & & \ldots & & a_n \\
   b_1 & a_1b_1\delta & a_2b_1\delta & & \ldots & & a_nb_1\delta \\
   b_2 & a_1b_2\delta & a_2b_2\delta & & \ldots & & a_nb_2\delta \\
   . & . & . & & \ldots & & . \\
   . & . & . & & \ldots & & . \\
   b_n & a_1b_n\delta & a_2b_n\delta & & \ldots & & a_nb_n\delta
\end{matrix}
\right].
\]
Observe that for every sub-determinant of \( A_n \) that is non-zero,
the corresponding sub-determinant of \( A_n(\delta) \) will also
remain non-zero. Thus \( \rk(A_n(\delta) ) = 2 \). But \(
\Rig(A_n(\delta),1) = 1 \) because if one changes \( \alpha \) to \(
\frac{1}{\delta} \) then every \( 2 \times 2 \) sub-determinant
becomes zero.

To summarize, we exhibited an infinite family $\{A_n\}$ of matrices
such that $\Rig(A_n, 1) = n-1$ and, given any $\epsilon_n > 0$, we
constructed an infinite family $\{A_n(\delta_n)\}$ such that
$A_n(\delta_n)$ is $\epsilon_n$-close to $A_n$ but
$\Rig(A_n(\delta_n),1) = 1$. This shows that the rigidity function is
in general not semicontinuous.

\subsubsection{Examples which are maximally rigid}
\label{subsubsec:maxfail}

The above example matrices are not maximally rigid. Might it be that
for matrices of highest rigidity, semicontinuity holds? We now produce
examples of matrices with maximum rigidity where the semi-continuity
property of rigidity fails.  Let
\[
A = \left[
\begin{matrix}
a & b & c \\
d & e & 0 \\
g & 0 & i	
\end{matrix}
\right],
\]
where \( a, b, \ldots, i \) are non-zero rational numbers.  Notice
that changing 4 entries (namely \(a,b,d,e\)) will be enough to bring
the rank down to 1.  It is easy to verify that changing $3$ entries
will not suffice for a general choice of $a, \ldots, i$. Thus, \(
\Rig(A,1) = 4 = (3-1 )^2 = (n-r)^2 \), with $n=3$ and $r=1$.

Let \( M \) be a generic matrix and let \( \pi \) be the diagonal
pattern of size \( 3 \) (represented by variables \( t_1, t_2, t_3
\)). Consider
\[
M+T_{\pi} = \left[
\begin{matrix}
a+t_1 &   b   & c     \\
d & e+t_2 & f     \\
g &  h    & i+t_3
\end{matrix}
\right].
\]
It can be checked that the elimination ideal for target rank $r=1$ is
generated by \( bfg - cdh \).  Note that \( A \) satisfies this
equation and thus it follows that \( A \in \overline{ \RIG (3, 1, 3,
  \pi ) } \).  This implies that any Zariski open neighborhood of \( A
\) intersects \( \RIG( 3, 1, 3, \pi ) \). This is a straightforward
consequence of the definitions. In fact, for any $\epsilon > 0$,
consider the matrix
$$
A(\delta) = \left[
\begin{matrix}
a & b & c \\
d & e & cd \delta \\
g & bg \delta & i	
\end{matrix}
\right],
$$ 
where $\delta \neq 0$ is chosen such that $\epsilon \geq \max \{cd
\delta, bg \delta \}$. Then $A(\delta)$ is within the open ball of
radius $\epsilon$ around $A$. Also, $\Rig(A(\delta),1) \leq 3$ because
we may change the diagonal entries to get the matrix
$$
B = \left[
\begin{matrix}
\delta^{-1} & b & c \\
d & bd \delta & cd \delta \\
g & bg \delta & cg \delta	
\end{matrix}
\right]
$$ 
which has rank $1$. Thus we have explicitly demonstrated that $A$
is in the Euclidean closure of $\RIG(3,1,3,\pi)$.

\subsection{Euclidean vs. Zariski Topology}
\label{sec:euclidean-zariski}

When defining semicontinuity, it is more natural to consider the
Euclidean topology. On the other hand, for algebraically defined
classes of matrices such as those in Section~\ref{sec:mainresult}, it
is more natural to study the Zariski closure. It is easy to see that
the Euclidean topology is in general finer than the Zariski topology,
i.e., closed sets in the latter are also closed in the
former. Interestingly, these two notions coincide in our context: we
show that the closures of the rigidity loci are equal in the Zariski
and Euclidean topology.
\begin{proposition}
The Euclidean Closure of \( \RIG(n, r, \leq k )(\C) \) equals its
Zariski Closure.
\end{proposition}
\begin{proof}
Recall that we can write \( \RIG(n, r, \leq k ) = \bigcup_{ \pi, ~ |
  \pi | = k } \RIG (n, r, \pi ) \). Thus, to prove the proposition, it
is sufficient to prove that for any pattern \( \pi \), the Euclidean
closure of \( \RIG(n, r, \pi ) \) equals its Zariski Closure.  By
Closure Theorem, there exists a subvariety \( V \) strictly contained
in \( \W := \overline{\RIG (n, r, \pi )} \) such that $\W ( \C) - V (
\C) \subseteq \RIG ( n, r, \pi ) ( \C) \subseteq \W ( \C)$.  Since \(
\W( \C) \) is closed in the Euclidean topology, we will be done if we
prove that the Euclidean closure of \( \W( \C) - V ( \C) \) is \( \W (
\C) \). This is precisely the statement of the following lemma from
\cite{Sha1}, which we state below for easy reference. Also note that,
by Corollary \ref{coro:decomp}, \( W \) is an irreducible variety for
every pattern \( \pi \) and hence the lemma is applicable.
\end{proof}
\begin{lemma}\emph{\bf{ (\cite[Lemma 1, page 124]{Sha1})}}
If \( X \) is an irreducible algebraic variety and \( Y \) a proper
subvariety of \( X \) then the set \( X ( \C ) - Y (\C) \) is dense in
\( X ( \C) \).
\end{lemma}

Let us consider the matrix $A$ in \eqref{eq:matrix}. We showed earlier
that $A \in \RIG(3,1,2)$ and yet there are matrices arbitrarily close
to it that belong to \( \RIG(3,1,1) \). Thus \( A \) is in the
Euclidean closure of \( \RIG(3,1,1) \), hence it is also in the
Zariski closure of \( \RIG(3,1,1) \). Let us verify this directly.

We want to check that $A \in
\W ( 3,1,\le 1)$.  We do this by showing a pattern $\pi$ such
that $A \in \W (3,1,\pi)$.  Let \( \pi := \{ (1,1) \} \).  Let
us write:
\[
X + t_1 := \left[
\begin{matrix}
x_1 + t_1 & x_2 & x_3 \\
x_3 & x_5 & x_6 \\
 x_7 & x_8 & x_9
\end{matrix}
 \right],
\]
where \( t_1 \) is the variable associate to \( \pi \). We obtain
\begin{eqnarray*}
I (3, 1, 1, \pi ) & = &  \langle t_1 x_5 + x_1 x_5 - x_2 x_4, t_1 x_6 + x_1 x_6 - x_3 x_4, \\
&		& ~~t_1 x_8 + x_1 x_8 - x_2  x_7, t_1 x_9 + x_1 x_9 - x_3 x_7, \\
&   & ~~x_2 x_6 - x_3 x_5,  x_2 x_9 - x_3  x_8, x_4 x_8 - x_5 x_7,  \\
&   & ~~x_4 x_9 - x_6 x_7, x_5 x_9 - x_6 x_8 \rangle.
\end{eqnarray*}
Eliminating $t_1$ from \( I (3, 1, 1, \pi ) \) using the Gr\"obner
Basis algorithm we get
\begin{eqnarray*}
EI (3, 1, 1, \pi ) & = & \langle x_2 x_6 - x_3 x_5, x_2 x_9 - x_3 x_8,
x_4 x_8 - x_5 x_7, \\ & & ~~x_4 x_9 - x_6 x_7, x_5 x_9 - x_6 x_8
\rangle.
\end{eqnarray*}
It is now easy to verify that \( A \) satisfies these generating
polynomials and hence $A \in \W (3,1,\pi)$.

\subsection{Some matrices with good neighborhoods}
\label{sec:good-nbhds}

Although the semicontinuity property fails for the rigidity function
over the entire space of matrices, we observe below that around
certain nice matrices the rigidity function does remain nondecreasing
within a small neighborhood.

In fact, the examples above suggest a technique for proving that there
is an $\epsilon$ such that the $\epsilon$-neighborhood of some
explicitly constructed matrix does not contain matrices of strictly
smaller rigidity. For this, we consider the Zariski closure of
matrices of rigidity at most $k-1$ (for some $k$). For a matrix $M$ of
rigidity at least $k$, if we prove that it does not lie in the above
closure, then it means that it is in the complement of a Zariski
closed set, and hence in a Euclidean open set. Thus there must be an
$\epsilon$ such that the $\epsilon$-neighborhood of $M$ does not
contain matrices of rigidity smaller than $k$.

We illustrate the above technique by an example: Consider the matrix
\[ M := \left[ \begin{matrix}
    2 & 3  & 5\\
    7 & 11 & 13\\
    17 & 19 & 23
  \end{matrix} \right] \in M_3(\C) .
\]
This is a matrix all of whose entries are distinct prime numbers. We
will show below that $M \in \RIG(3,1,4)$, but $M \notin \W (3,1, 3)$.

We will prove this by ruling out all possible patterns $\pi$ of size
3. We can quickly rule out some of these patterns as follows. Consider
the pattern matrix $T_\pi$ such that
\[
M + T_\pi = \left[
  \begin{matrix}
    a+t_1 & b+t_2 & c+t_3 \\
    d & e & f \\
    g & h & i
  \end{matrix}
\right].
\]
Then the equation $ \left| \begin{matrix} e & f \\ h & i
  \\ \end{matrix} \right| = 0 $ belongs to the associated elimination
ideal. Note here that the matrix $M$, due to its choice of entries,
has the property that all the submatrices have full rank. Hence the
above equation is obviously not satisfied by $M$. Similarly, we can
rule out patterns \( \pi \) of size 3 for which either any row or any
column contains at least two non-zero entries.  Thus, to prove the
claim we need to only rule out patterns $T_\pi$ that touch all \( 2
\times 2 \) minors. Thus, up to permutations (since choice of primes
in $M$ could be arbitrary but distinct) we need to check the case when
$T_\pi$ has the variables on the diagonal:
\[
M+T_\pi = \left[
  \begin{matrix}
    a+t_1 &   b   & c     \\
    d & e+t_2 & f     \\
    g & h & i+t_3
  \end{matrix}
\right].
\]
In this case, the elimination ideal is generated by a single
polynomial, namely $ bfg-cdh $, which again $M$ does not satisfy.
Since up to permutations, all patterns of size \( 3 \) can be written
as one of the above, we conclude that $M \notin \W(3,1,3)$. In
addition, by the argument outlined earlier, this also implies that for
the matrix $M$, there is an $\epsilon$ such that all the matrices in
the $\epsilon$-neighborhood are outside $\W(3,1,3)$.

Note that for the purposes of this argument, we can get by with much
less: instead of populating the matrix with distinct primes, we could
take a Vandermonde matrix
$$
\left[
\begin{matrix}
    1 & p  & p^2\\
    1 & q & q^2\\
    1 & r & r^2
\end{matrix}
\right],
$$
where $p,q,r$ are distinct primes.

\section{Conclusions and Further Research}
\label{sec:conc}

In this paper, we considered the problem of finding $n \times n$
matrices of highest possible rigidity, i.e. $(n-r)^2$, for target rank
$r$. In the first part, we presented a proof in the language of
algebraic geometry, of Valiant's classical theorem that most matrices
over $\C$ have rigidity exactly $(n-r)^2$. In addition, we are able to
compute the exact dimension of the variety of matrices of rigidity
strictly less than $(n-r)^2$. A natural question is to ask for the
degrees and other geometrical properties of the loci
$\overline{\Rig(n,r,\leq k)}$ of matrices with rigidity at most $k$
(we computed the dimensions in Theorem
\ref{thm:dim-of-rigid-matrices}).

Our second and main contribution is to construct certain explicit
matrices of highest possible rigidity over $\C$. Entries of these
matrices are primitive roots of unity of orders approximately
$\exp(n^2 \log n)$. While these matrices have a concrete and succinct
algebraic description, they are still not explicit from a
computational complexity perspective. In particular, the main open
question of constructing polynomial time computable matrices of even
superlinear rigidity is still wide open.

It is unclear whether the exponential orders, $\exp(n^2 \log n)$, for
the roots of unity used in the matrices of
Theorem~\ref{thm:rigid-matrices-over-C} are necessary. It would be
interesting to obtain matrices of \emph{optimal} rigidity using roots
of polynomial, or even $\exp(n)$, orders. Results on effective
Nullstellensatz used in the proof of
Theorem~\ref{thm:degree-bound-on-elim-ideals} show exponential degree
bounds for polynomials in elimination ideals are in general
unavoidable. Thus any improvements may have to exploit the special
nature of the elimination ideals of matrices of rigidity less than
$(n-r)^2$. In particular, as remarked in earlier sections, elimination
ideals of determinantal varieties are objects worthy of study in this
context. Note that \cite{Lok06} constructs matrices of
\emph{asymptotically optimal} rigidity using roots of unity of
polynomial orders, using different and more elementary arguments.

Both our lower bound and the one from \cite{Lok06} rely on the fact
that the corresponding matrices live in number fields of at least
exponentially large dimensions. This dimension can be viewed as an
algebraic measure of explicitness of the matrix; the lower the
dimension, the more explicit the matrix. Constructing matrices of high
rigidity whose entries come from number fields of polynomial dimension
is an open question.

A particularly interesting problem is whether a Vandermonde matrix $V
= \left(x_i^{j-1}\right)_{ij}$ with algebraically independent
coordinates $\{x_i\}$ has maximal rigidity. To analyze this question,
one would look at the rigidity loci restricted to the subvariety of
Vandermonde matrices. If this question has an affirmative answer, we
believe that one may proceed using the Nullstellensatz (as we have
done in here) to construct explicit Vandermonde matrices with entries
being algebraic numbers, of significantly smaller complexity than
those in this paper. We mention in passing that the finite Fourier
transform matrix $\mathcal{F} = (\zeta_n^{(i-1)(j-1)})_{ij}$, which is
a Vandermonde matrix, does \emph{not} have maximal rigidity (for
instance, for target rank $\lfloor 3n/4 \rfloor$, as long as $n >
16$).

In the final part of the paper, we try to understand the topological
behavior of the rigidity function in the neighborhood of highly rigid
matrices. Our main motivation for this line of investigation comes
from the intuition that we may be able to find sufficiently explicit
rational matrices of (moderately) high rigidity that approximate the
complex matrices of (very) high rigidity that seem easier to find.  We
give examples to show that the rigidity function is in general not
semi-continuous, meaning that within a small (Zariski or Euclidean)
neighborhood of certain matrices, the rigidity function can strictly
decrease. On the other hand, around many ``natural and interesting''
matrices, we find that the rigidity function is actually nondecreasing
within a small neighborhood. We think that a better understanding of
the topology of the stratification of $M_n(\C)$ by the subsets
$\Rig(n,r,k)$ will have a bearing on the complexity-theoretic problem
of constructing matrices of high rigidity.

\bibliographystyle{alpha}
\bibliography{elimination-rigidity}

\appendix

\section{Background on Algebraic Geometry}
\label{sec:prelims}

In this section, we recall some basic notions from algebraic
geometry. Much of this background can be found in \cite{HS00} and
\cite{EH00}.

We aim for a relatively elementary description: in particular, we will
identify a variety with the set of its points over the algebraic
closure, rather than thinking of its points as the prime ideals of a
ring (the scheme-theoretic point of view).

Let \( F \) be a field. Let \( \Fbar \) denote a fixed algebraic
closure of \( F \). Let \( x_1 , \cdots, x_n \) be \( n \)
algebraically independent variables over \( F \). Let \( F [ x_1 ,
  \cdots , x_n ] \) be the polynomial ring in \( n \) variables over
\( F \). An ideal \( I \) is by definition a sub-module of the ring \(
F [ x_1 , \cdots , x_n ] \). More explicitly, \( I \) is a subset of
\( F [ x_1 , \cdots , x_n ]\) which is a subgroup of \( F [ x_1 ,
  \cdots , x_n ] \) under addition, and which is also closed under
multiplication by elements of \( F [ x_1 , \cdots , x_n ] \). The
ideal $I$ is {\em prime} if whenever $rs \in I$ with \( r,s \in F[x_1,
  \cdots, x_n] \), either $r \in I$ or $s \in I$.

An {\em affine algebraic variety} $S \subset \Fbar^n$ is a subset
$$ 
V(\Sigma) = \{ (a_1, \dots, a_n) \in \Fbar^n  :  f(a_1, \dots,
a_n) = 0 \textrm{ for all } f \in \Sigma \}
$$
for some subset $\Sigma$ of $ \Fbar[x_1, \ldots x_n] $. In
particular, $\Sigma$ may consist of polynomials with coefficients in $F$,
in which case we say that $V(\Sigma)$ is defined over $F$. In
particular, we have {\em affine $n$-space} $\Af^n = V(\{0\})$, and any
affine algebraic variety is a subset of some $\Af^n$ cut out by a set
of polynomials.

If $I_{\Sigma}$ is the ideal generated by $\Sigma$ in $\Fbar[x_1,
\dots, x_n]$ (or in $F[x_1, \dots, x_n]$ if $\Sigma \subset F[x_1,
\dots, x_n]$), it is clear $V(I_\Sigma) = V(\Sigma)$. Therefore we may
restrict attention to zero sets of ideals from now on. The Hilbert
Basis theorem says that every ideal of a polynomial ring over a field
is finitely generated, so we observe that we could always have started
with a finite set of generators $\Sigma$. Since each generator is a
polynomial with finitely many coefficients, it follows that any
algebraic variety $V(I)$ may be defined over some finite extension of
$F$.

For an affine variety $V(I)$ and an extension \( L \) of \( F \), we
define its $L$-rational points to be
$$ 
V(I)(L) := \{ (a_1, \dots, a_n) \in L^n  :  f(a_1, \dots, a_n)
= 0 \textrm{ for all } f \in I \}.
$$

The algebraic variety $V(I)$ is a geometric object with a natural
structure of a topological space, where the closed subsets are $V(J)$
for ideals $J \subseteq \Fbar[x_1,\ldots,x_n]$ containing $I$. This is
called the \emph{Zariski topology}.

On the other hand, given a subset \( S \) of \( \Fbar^{\, n} \), let
us define \( I(S) \) to be the set of polynomials \( f \in \Fbar [
x_1 , \cdots , x_n ] \) such that \( f ( s) = 0~ \forall s \in S \);
it follows that \( I(S) \) is an ideal of \( \Fbar [ x_1 , \cdots ,
x_n ] \). If $S \subset F^n$, then it is not hard to see that one can
choose generators of $I(S)$ to lie in $F[x_1, \dots, x_n]$.  We can
then associate the ideal $I_F(S) = I(S) \cap F[x_1, \dots, x_n]$ to
$S$. Note that $I_F(S) \cdot \Fbar[x_1, \dots, x_n] = I(S)$.

For any ideal $I \subset \Fbar[x_1, \dots, x_n]$, let us define
\[
\sqrt{I} := \{ f \in \Fbar [ x_1 , \cdots , x_n ] : \exists m \in
\N ~\mathrm{such~that}~f^m \in I \} .
\]
\( \sqrt{I}  \) is called the \emph{radical} of the ideal \( I \).
We then have the following fundamental theorem.
\begin{theorem}(Hilbert's Nullstellensatz)
For an ideal \( I \) of \( \Fbar [ x_1 , \cdots , x_n ] \), \(
\sqrt{I} = I ( V(I) ) \).
\end{theorem}

We will always deal with radical ideals, namely those $I$ which are
equal to \(\sqrt{I} \).

Given a subset \( S \) of \( \Fbar^n \), the Zariski-closure of \( S
\), denoted by \( \overline{S} \), is the smallest \emph{algebraic}
variety of \( \Fbar^n \) containing \( S \). In other words, we have
$\overline{S} = V(I(S))$.

We say that an algebraic variety \( X \) is {\em irreducible} if it
can not be written as a union of two algebraic varieties \( X_1 \) and
\( X_2 \) properly contained in $X$.  Note that $X$ is irreducible if
and only if $I(X)$ is a prime ideal.

A morphism \( \phi:X \subseteq \Af^n \to \Af^1 \) from an affine
closed subvariety of affine $n$-space to the affine line is a
polynomial map \( (x_1, \ldots x_n) \mapsto p(x_1, \ldots, x_n) \)
where \( p \) is a polynomial. We naturally extend this to a morphism
between affine varieties.
\begin{definition}
Let \( X \subseteq \Af^n \) and \( Y \subseteq \Af^m \) be two
closed affine varieties. A morphism \( \phi : X \to Y \) is defined to be a map
\( \phi \) whose components are polynomials. In other words,
$\phi$ has the form:
\[
\phi(x_1, \ldots x_n) = (f_1(x_1, \ldots, x_n), \ldots, f_m(x_1, \ldots, x_m))
\] where \( f_1, \ldots f_m \) are polynomials, and with
the property that it maps the subset \( X \) to \( Y \).

The morphism $\phi$ is called \emph{dominant} if $\phi(X)$  is dense in $Y$.
\end{definition}

Let $X = V(I) \subset \Af^n$ be an affine algebraic variety, where $I
\subset \Fbar[x_1, \dots, x_n]$, and let \( \Fbar(X) \) denote the
ring of fractions of the quotient ring \( R = \Fbar[x_1, \ldots,
x_n]/I(X) \). If \( I(X) \) is a prime ideal, \( \Fbar(X) \) is a field and
is called the {\em function field} of \( X \).  Elements of the
function field $\Fbar(X)$ are called the set of {\em rational functions}
on the variety \( X \).

\begin{definition}
  Let \( K \) be a finitely generated extension field over a base
  field \( F \). Let \( T \) be a maximal set of algebraically
  independent elements of \( K \) over \( F \). Such a \( T \) is
  called a \emph{transcendence basis} of \( K \) over \( F \). It can
  be proved that the cardinality \( | T | \) is independent of \( T
  \), and is called the \emph{transcendence degree} of \( K \) over \(
  F \) and will be denoted by \( \trdeg ( K /F) \).
\end{definition}
\begin{definition}
  The dimension of an irreducible affine variety \( X \subseteq
  \Fbar^n \), denoted by \( \dim (X) \), is the transcendence degree
  of the function field \( \Fbar(X) \) of the variety \( X \) over the
  base field $\Fbar$. Thus, \( \dim(X) := \trdeg ( \Fbar(X) / \Fbar )
  \).
\end{definition}
For easy reference we state a lemma below that is an immediate consequence of Theorem 4.4, Chapter 1, of \cite{Hart77}.
\begin{lemma}
\label{lem:dim-image-rational-isomorphism}
Let \( \phi: X \to Y \) be a dominant morphism of irreducible
varieties over $F$. Then \( \phi \) induces a natural embedding \(
\phi^{\ast} : \Fbar(Y) \hookrightarrow \Fbar(X) \). In particular,
\[ \dim (Y) = \trdeg ( \Fbar(Y)/ \Fbar) \leq \trdeg (\Fbar(X)/\Fbar) = \dim(X). \]
\end{lemma}
If a variety $X$ is not irreducible, we define its dimension to be the
maximum of the dimensions of its (finitely many) irreducible
components. The conclusion of
Lemma~\ref{lem:dim-image-rational-isomorphism} that $\dim(Y) \leq
\dim(X)$ continues to hold for a dominant morphism $X \rightarrow Y$
of varieties which may be reducible.

We have described closed affine subvarieties of affine $n$-space. In
particular, a closed subset of $\Af^n$ that is defined by a single
polynomial $f$ in $n$ variables is called a hypersurface $V(f)$. Now,
it can be shown that the Zariski topology of $\Af^n$ has a basis of
open sets given by the complements of these hypersurfaces, $D(f) =
\Af^n \backslash V(f)$. In fact, $D(f)$ is itself isomorphic to an
affine variety, namely the hypersurface $fy = 1$ in $\Af^n \times
\Af^1_y$. In general, a space which we can thus identify naturally
with a closed affine subvariety in some affine space (in a sense that
we will not make precise here) is called an affine variety. An
important example of this is the open subset \( GL_n = D (\det ) = M_n
\backslash V(\det) \) of invertible matrices in \( M_n \), where \(
\det \) stands for the determinant polynomial.

A general algebraic variety $X$ is obtained by glueing together
various pieces $X_i$ such that $X_i$ is an affine variety. The notion
of gluing means that there are open varieties $U_{ij} \subset X_i$
and compatible isomorphisms $U_{ij} \rightarrow U_{ji}$ between them
(so that we can think of $U_{ij}$ as the intersection of $X_i$ and
$X_j$).

\end{document}